\documentclass[prl,aps,longbibliography,twocolumn,superscriptaddress]{revtex4}

\usepackage{graphicx}
\usepackage{times}
\usepackage{color}
\usepackage{float}
\usepackage{amssymb,amsmath}
\usepackage{epsfig}
\usepackage{epstopdf}
\usepackage{hyperref}
\usepackage{commath}
\usepackage{color,soul}

\begin{document}

\title{Synchronization of chaotic systems: a microscopic description}

\author{Nir Lahav}\thanks{Corresponding author: freenl@gmail.com}
\affiliation{Department of Physics, Bar-Ilan University, 52900 Ramat Gan, Israel}

\author{Irene Sendi\~na-Nadal}
\affiliation{Complex Systems Group {\& GISC}, Universidad  Rey Juan Carlos, 28933 M\'ostoles, Madrid, Spain}
\affiliation{Center for Biomedical Technology, Universidad Polit\'ecnica de Madrid, 28223 Pozuelo de Alarc\'on, Madrid, Spain}

\author{Chittaranjan Hens}
\affiliation{Department of Mathematics, Bar-Ilan University, 52900 Ramat Gan, Israel}

\author{Baruch Ksherim}
\affiliation{Department of Mathematics, Bar-Ilan University, 52900 Ramat Gan, Israel}

\author{Baruch Barzel}
\affiliation{Department of Mathematics, Bar-Ilan University, 52900 Ramat Gan, Israel}

\author{Reuven Cohen}
\affiliation{Department of Mathematics, Bar-Ilan University, 52900 Ramat Gan, Israel}

\author{Stefano Boccaletti}
\affiliation{CNR-Institute of complex systems, Via Madonna del Piano 10, 50019 Sesto Fiorentino, Italy}
\affiliation{Unmanned Systems Research Institute, Northwestern Polytechnical University, Xi’an 710072, China}

\begin{abstract}

The synchronization of coupled chaotic systems represents a fundamental example of self organization and collective behavior. This well-studied phenomenon is classically characterized in terms of macroscopic parameters, such as Lyapunov exponents, that help predict the system's transitions into globally organized states. However, the local, microscopic, description of this emergent process continues to elude us. Here we show that at the microscopic level, synchronization is captured through a gradual process of \textit{topological adjustment} in phase space, in which the strange attractors of the two coupled systems continuously converge, taking similar form, until complete \textit{topological synchronization} ensues. We observe the local nucleation of topological synchronization in specific regions of the system's attractor, providing early signals of synchrony, that appear significantly before the onset of complete synchronization. This local synchronization initiates at the regions of the attractor characterized by lower expansion rates, in which the chaotic trajectories are least sensitive to slight changes in initial conditions. Our findings offer a fresh and novel description of synchronization in chaotic systems, exposing its local embryonic stages that are overlooked by the currently established global analysis. Such local topological synchronization enables the identification of configurations where prediction of the state of one system is possible from measurements on that of the other, even in the absence of global synchronization.

\end{abstract}

\maketitle

Synchronization underlies numerous collective phenomena observed in nature \cite{Pikovsky2001}, providing a scaffold for emergent behaviors, ranging from the acoustic unison of cricket choruses to the coordinated choreography of starling flocks \cite{Glass1988}. Synchronization phenomena have also found applications pertaining to human cognition, providing a theoretical framework to study insomnia, epilepsy and Parkinson's disease \cite{Ulhaas2006}, as well as perception, memory and consciousness \cite{Rodriguez1999,Klimesch1996,Singer2011}.

Special attention has been given to the synchronization of chaotic systems \cite{pecora1990}, which, due to their high sensitivity to initial conditions, intrinsically defy synchrony. Therefore, characterizing and understanding the transition from incoherence to synchrony in such systems is of fundamental importance \cite{Pikovsky2001}. The phenomenon is often observed by tracking the coordinated behavior of two slightly mismatched coupled chaotic systems, namely two systems featuring a minor shift in one of their parameters. As the coupling strength increases, a sequence of transitions occurs, beginning with no synchronization, advancing to phase synchronization \cite{jugphase}, lag synchronization \cite{bocavalladares}, and eventually, under sufficiently strong coupling, reaching complete synchronization. These transitions have been thoroughly characterized by means of \textit{global} indicators, \textit{i.e.} by monitoring sign changes in the Lyapunov spectrum \cite{jugphase} or identifying crossover points between different scaling laws \cite{boccaexp}. A more detailed view can be attained by focusing on the properties of unstable periodic orbits embedded in the attractor \cite{Pikovsky1997,Pazo2003,Yanchuk2003,Cvitanovic1991}. Such orbits, constituting long trajectories within the attractors, provide crucial indicators for different types of synchronization, helping charaterize the stability of the synchronization manifold and playing a crucial role also in the process of desynchronizaiton \cite{Yanchuk2003}. 

Hence, synchronization is characterized at the macroscopic level through the Lyapunov spectrum and at the meso-scopic level through the non-localized unstable periodic orbits. We offer to complement these views by developing a local description of synchronization as a continuous process of \textit{topological adjustment} in phase space between two different strange attractors. As the coupling strength increases, we show that the structures of the attractors of the two coupled systems begin to assimilate, until, at the point where complete synchronization is attained, they reach perfect topological matching. In order to demonstrate that, we expose the synchronization focal points, a concept that has been previously suggested \cite{boccapeco}, and identify the precise nucleation points in phase space where synchronization is locally established. Our analysis allows us to uncover the microscopic mechanisms underlying the synchronization transition and helps us tracks the gradual microscopic onset of synchronization, preceding the macroscopic transition to complete synchrony. 


We demonstrate our findings on one of the fundamental examples in the context of synchronization, capturing two slightly mismatched chaotic R\"{o}ssler oscillators \cite{roessler} coupled in a master-slave configuration. Focusing on this commonly studied system allows us to put our findings in context, and directly compare them with the state-of-the-art pertaining to chaotic synchronization. 
The equations of motion driving these oscillators take the form

\vspace{-5mm}
\begin{align}
\nonumber
\dot{\bf x}_1  &= f_{1}({\bf x}_1)
\\
\dot{\bf x}_2 &= f_{2}({\bf x}_2) +k ({\bf x}_1-{\bf x}_2),
\label{ecuaciones}
\end{align}
\noindent
where ${\bf x}_1 \equiv (x_1,y_1,z_1)$ and ${\bf x}_2 \equiv (x_2,y_2,z_2)$ are the vector states of the master and slave oscillators respectively, $k$ is the coupling strength and $f_{1,2}({\bf x})= (-y - z, x  + ay, b + z(x - c_{1,2}))$.
Without loss of generality we set the parameters to $a = 0.1$ and $b = 0.1$ identically across the two oscillators, and express the slight mismatch between the master and the salve through the parameters $c_1 = 18.0$ vs.\ $c_2 = 18.5$. The system (\ref{ecuaciones}) describes a unidirectional master (${\bf x}_1$) slave (${\bf x}_2$) form of coupling, uniformly applied to all coordinates $x,y$ and $z$. Under this directional coupling scheme we can track and quantify the process of synchronization in a controlled fashion, as the slave gradually emulates the behavior of the master, while the master continues its undisturbed oscillations. The uniform coupling, under which synchronization is entered, but never exited, allows another layer of observation simplicity, which we relax later on, by investigating single variable coupling, observing a richer set of synchronization transitions.

As the coupling $k$ in (\ref{ecuaciones}) is increased, the system (1) undergoes a set of transitions, first from incoherence to phase synchronization \cite{jugphase}, and then from phase to complete synchronization. The first transition is characterized by the phase order parameter 
\begin{equation}
r = \lim_{\tau \to \infty} \left[ \frac{1}{\tau} \int_{0}^{\tau} \frac{1}{2}
\big| e^{i\theta _1} + e^{i\theta _2} \big| \dif t \right],
\label{r}
\end{equation}
\noindent
where $\theta_{1,2}\equiv \arctan (\frac{y_{1,2}}{x_{1,2}})$, which approaches unity if both oscillators are in phase, \textit{i.e.} $\theta_1 = \theta_2$. The second transition occurs when the synchronization error 
\begin{equation}
E = \lim_{\tau \to \infty} \left[ \frac{1}{\tau} \int_{0}^{\tau} \|{\bf x}_1-{\bf x}_2\| \dif t \right]
\label{E}
\end{equation}
\noindent
vanishes, indicating that ${\bf x}_1$ and ${\bf x}_2$ oscillate in perfect unison \cite{notilla}.

\begin{figure}
\centering
\includegraphics[width=0.65\columnwidth]{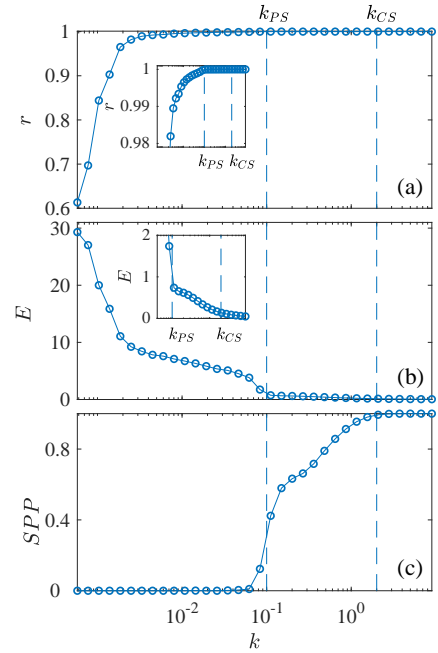}
\vspace{-4mm}
\caption[]{(Color online). \textbf{Microscopic build-up of synchronization}. (a) Phase order parameter $r$ vs.\ $k$ as obtained from the coupled R\"{o}ssler oscillators (\ref{ecuaciones}). Phase synchronization ($r \rightarrow 1$) occurs for $k > k_{\rm PS} \sim 0.1$ (vertical dashed line). The inset shows the behavior of $r$ in the vicinity of the transition. (b) Synchronization error $E$ vs.\ the coupling strength $k$. Under complete synchronization we have $E \rightarrow 0$, obtained for $k \ge k_{\rm CS} \sim 2.0$ (vertical dashed line, see also inset). (c) SPP vs.\ $k$ allows us to observe the microscopic path to synchronization. As expected, SPP $\rightarrow 1$ in complete synchronization ($k \ge k_{\rm CS}$), however the local synchronization points appear much earlier. For instance, SPP $\sim 30\%$ already at $k \sim 0.1$, indicating that local synchronization, at selected focal points of the coupled strange attractors, is already present.
\label{fig1}}
\vspace{-3mm}
\end{figure}

These two transitions are observed in Fig.\ \ref{fig1}a,b, where we present the order parameters $r$ and $E$ vs.\ the coupling $k$. Phase synchronization emerges as $r \rightarrow 1$ at $k_{\rm PS} \sim 0.1$, and complete synchronization follows as $E \rightarrow 0$ at $k_{\rm CS} \sim 2.0$ (vertical dashed lines). The common approach for observing these transitions is to track sign changes in the system's Lyapunov exponents, as obtained from ensemble averages of the eigenvalues of (1)'s Jacobian matrix

\begin{equation}
  \label{jacobian}
J\equiv
\begin{pmatrix}
  0&-1&-1& 0&0&0\\
 1&a&0&0&0&0  \\
z_1&0&x_1-c_1&0&0&0\\
 k&0&0&-k&-1&-1 \\
0&k&0&1& a-k&0\\
0&0&k& z_2&0& x_2-c_2-k
\end{pmatrix}.
\end{equation}
\noindent
Averaging $J$'s eigenvalues over all phase space configurations spanned by the chaotic trajectory, one obtains the system's six Lyapunov exponents, which help characterize the state of the system. In the disordered regime ($0 \leq k < k_{\rm PS}$) the Lyapunov spectrum comprises two positive, two vanishing and two negative exponents; under phase synchronization ($k_{\rm PS} \le k < k_{\rm CS}$) it transitions into two positive, one zero and three negative exponents; and at $k \ge k_{\rm CS}$ one of the two positive exponents also becomes negative, marking the onset of complete synchronization.

Obtained by ensemble averages, the Lyapunov spectrum is a global indicator of the state of the system, aggregating over the behavior of the entire system trajectory in phase space. It provides, however, little insight into the specific microscopic states of the system along this trajectory. Hence, to complement this global perspective, we wish to inspect the local emergence of synchronization by assessing the level of coherence of different points in the two oscillators' phase space. To observe this we measure the synchronization points percentage (SPP) index \cite{boccapeco,Pastur04}, which quantifies the fraction of points in phase space for which there exists a local continuous surjective function to map the state of the master with that of the slave. If, within some area in the vicinity of these points such a continuous mapping exists, one can predict the specific state of the slave system directly from measuring the master state, representing a local topological coherence of the master-slave duo \cite{Pecora95}. Thus, the SPP index can monitor the gradual passage from local to global synchronization in the system. 

\begin{figure}
\centering
\includegraphics[width=0.8\columnwidth]{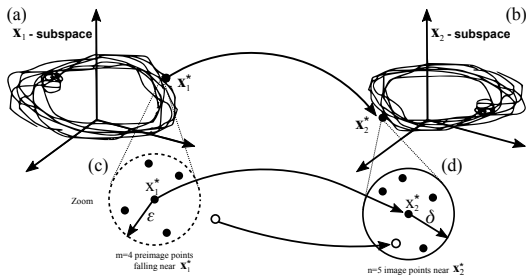}
\caption[]{
\textbf{Calculating the synchronization points percentage (SPP) index.} 
(a) Trajectory in the ${\bf x}_1$ sub-space, capturing the master attractor. (b) Matching trajectory in the ${\bf x}_2$ sub-space, \textit{i.e.} the slave attractor. Each include $N = 10^4$ points at fixed time intervals. 
(c) - (d) We select two points ${\bf x}_1^*$ and its image ${\bf x}_2^*$, together with a small environment, or neighborhood, within radii $\epsilon$ (${\bf x}_1^*$) and $\delta$ (${\bf x}_2^*$) for each point. In (c) we mark the $m < n$ pre-image points surrounding ${\bf x}_1^*$ (black) - in this case only $4$ of the five points correspond to ${\bf x}_2^*$'s environment. (d) $n = 5$ closest points of ${\bf x}_2^*$. Not all pre-image points are guaranteed to fall within ${\bf x}_1^*$’s neighborhood, hence $m < n$. For example, see the hollow circle in ${\bf x}_2^*$’s environment, whose pre-image falls outside that of ${\bf x}_1^*$. SPP quantifies the level of correspondence between all such environments to estimate the likelihood of a continuous mapping between the attractors.}
\label{fig2}
\end{figure}


In order to calculate SPP one has first to identify a proper domain and a proper co-domain for the statistical search of the existence of functional dependencies. We start by considering a set of $N = 10^4$ points, spread uniformly in time, in the ${\bf x}_1$-subspace of the phase space, and the corresponding $N$ images in the ${\bf x}_2$-subspace. Next, we select a specific pair of points, ${\bf x}_1^*$ in ${\bf x}_1$ and its image ${\bf x}_2^*$ in ${\bf x}_2$. We then consider the $n = 5$ closest points to ${\bf x}_2^*$, \textit{i.e.} ${\bf x}_2^*$'s neighborhood, and identify the $m$ pre-image points falling within a similar size neighborhood of ${\bf x}_1^*$. Here $ m \leq n$, as the neighborhood of ${\bf x}_2^*$ might also host images of points not falling within the close neighborhood of ${\bf x}_1^*$ (Fig.\ \ref{fig2}). If indeed neighborhoods in ${\bf x}_2$ tend to correspond to pre-image neighborhoods in ${\bf x}_1$, \textit{i.e.} $m$ tends to be larger than expected by chance, it is likely that a continuous function between the attractors is, indeed, at play.

To quantitatively assess our degree of confidence that such a local continuous function underlies the observed correspondence we estimate the probability for the observed mapping to emerge at random. The probability of a single point falling within the neighborhood of ${\bf x}_2^*$ at random is $n/N$, and therefore the probability for the $m$ points around ${\bf x}_1^*$ to have, by pure chance, images inside the neighborhood of ${\bf x}_2^*$ is $p_m=(n/N)^m$. We compare this probability to 

\begin{equation}
b_p = \max_{q=1,...,m} \big[ B(q,m; P) \big],
\label{bp}
\end{equation}

\noindent
where $B(q,m; P)$ is the binomial distribution, providing the probability that $q \leq m$ events out of $m$ attempts are realized for a process of elementary probability $P$. Setting $P=n/N$, $b_p$ represents the maximum over $q$ of the probability that, out of a given $m$ points, $q$ will fall into the neighborhood of ${\bf x}_2^*$. Hence, we estimate the level of confidence for the existence of a continuous mapping from ${\bf x}_1$ to ${\bf x}_2$ through the ratio 

\begin{equation}
\gamma = \dfrac{p_m}{b_p},
\label{theta}
\end{equation}

\noindent
which approached zero in the limit where $b_p \gg p_m$, indicating a strong likelihood for the existence of a local continuous (at least surjective) function mapping states in the vicinity of ${\bf x}_1^*$ into states around ${\bf x}_2^*$ \cite{Pastur04}. In our calculations we set a threshold of $\gamma = 0.1$ 
to discern whether or not two local neighborhoods are linked through a functional relationship, namely a pair $({\bf x}_1^*,{\bf x}_2^*)$ is rendered synchronized if its neighborhoods scored $\gamma < 0.1$. Denoting the number of synchronized pairs by $n^*$, we define the SPP index as the fraction of locally synchronized points in the system $n^*/N$. In the limit SPP$\to 1$ there exists a unique, global, continuous function from one subsystem to the other \cite{Pastur04}, and hence the slave subsystem is fully predictable from measurements of the master oscillator.



In Fig.\ \ref{fig1}c we show the SPP index vs.\ $k$, which provides us with a novel perspective into the microscopic process underlying the synchronization transition. It shows that SPP gradually increases with $k$, with a sizable portion of points in phase space undergoing local synchronization already at low coupling, much before the emergence of complete global synchronization. For example, already around $k \sim k_{\rm PS}$, significantly before the onset of complete synchronization, we observe SPP $\sim 0.3$, indicating that for a significant part of the slave trajectory, its state already can be predicted from that of the master's. This is despite the fact that complete global synchronization will only be observed around $k = k_{\rm CS} \sim 2.0$. This finding concurs with previous works focusing on bidirectional coupling \cite{boccapeco}, showing that, even when lacking global amplitude correlations, phase synchronization implies predictability of one system's state from that of the other for a rather large portion of the phase space. Such predictability, or mapping, between the instantaneous states of the two coupled systems, is precisely captured by the large SPP index observed as the system enters the phase synchronization regime. We emphasize, however, that this is not meant to suggest that SPP is an identifier of phase synchronization - indeed, $r$ (\ref{r}) is a suitable global parameter for that cause. Rather, the fact that already around the phase-synchronization transition SPP assumes a non-vanishing value, presents its power as a microscopic quantifier preceding complete synchronization, indicating that early traces of local synchronization appear much before than the full transition to complete synchrony. 

The gradual growth of SPP with $k$ can be understood as a continuous process of \textit{Topological Synchronization}: originally, at $k = 0$, the uncoupled master and slave attractors are topologically distinct, then, as $k$ is increased, the slave's strange attractor gradually assimilates to that of the master. Indeed, the SPP index captures the extent to which the two attractors can be mapped to each other, quantifying a level of topological similarity, which culminates in complete synchronization, where the two attractors become identical. Hence, complete synchronization, a transition occurring at $k_{\rm CS}$, begins at significantly lower coupling through a continuous microscopic process of topological synchronization, in which the slave and master attractors gradually converge until their trajectories become uniform. 

\begin{figure}
\centering
\includegraphics[width=0.8\columnwidth]{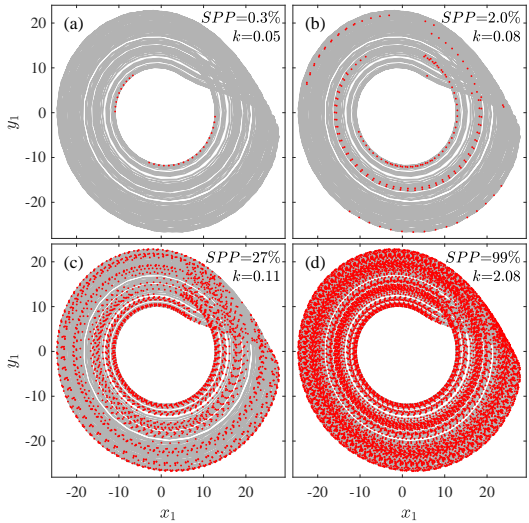}
\caption[]{(Color online). 
\textbf{Focal points of synchronization}. Synchronization points (red) in the $(x_1,y_1)$ phase plane of the master's strange attractor (grey), under increasing values of the coupling $k$. In each panel the SPP index (percentage) is also shown.}
\label{fig3}
\end{figure}

Next, we examine the characteristics of the initial \textit{synchronization points}, seeking whether they are randomly distributed within the master attractor, or restricted to specific areas in phase space. To observe this, in Fig.\ \ref{fig3} we plot the master attractor in the $(x_1,y_1)$ phase plane, marking the points of synchronization, \textit{i.e.} the points where the master state can be mapped to the slave state, by red dots. These red dots represent precisely the desired synchronization points, being the areas in the attractor that contribute to the SPP index. Interestingly, the synchronization points are confined to specific areas in the attractor, indicating that topological synchronization nucleates around preferred regions in the $(x_1,y_1)$ projection of the phase space. For example, under low coupling of $k = 0.05$, we have SSP $=0.3\%$ (Fig.\ \ref{fig3}a), an embryonic stage of topological synchronization. The crucial point in the current context is, however, that these selected $0.3\%$ synchronization points are distinctively located along specific areas in the attractor. This pattern continues as $k$ is increased in Fig.\ \ref{fig3}b-d, and the SPP index rises: indeed, the density of red points increases accordingly, yet, most importantly, they continue to cover only specific regions within the $(x_1,y_1)$ phase plane. Hence, topological synchronization arises from distinctive attractor focal points.

\begin{figure}
\centering
\includegraphics[width=\columnwidth]{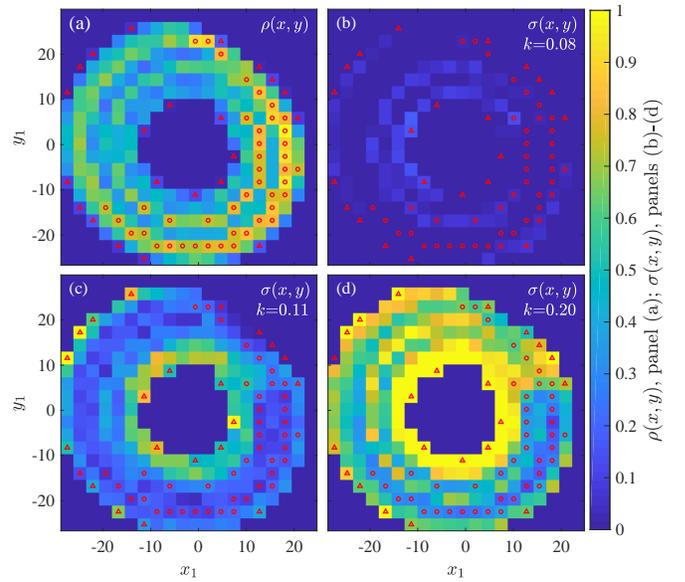}
\caption[]{(Color online). 
\textbf{Synchronization points and attractor density}. 
(a) The density $\rho(x,y)$ of points in the $(x_1,y_1)$ projection of the master attractor. We used a partitioning into $20 \times 20$ boxes and measured the density in each box ($x,y = 1,\dots,20$). Boxes with high density (above $70\%$ of $\max (\rho(x,y))$) are marked by red circles, while boxes with low density (below $20\%$ of $\max(\rho(x,y))$) are marked by red triangles. 
(b) - (d) The normalized local SPP $\sigma(x,y)$ under increasing values of the coupling strength $k$. Initial synchronization points are captured by the boxes with increased levels of $\sigma(x,y)$ (tending to yellow, common colorbar on the right). These early traces of synchronization favor the low density regions of the attractor, as observed by $\sigma(x,y)$'s consistent avoidance of the red circles vs.\ its tendency towards the red triangles.}
\label{fig4}
\end{figure}

To gain further insight into the microscopic onset of synchronization we investigate the connection between the synchronization points and the local density of the master attractor. First, we divide the $(x_1,y_1)$ plane into $l \times l$ small boxes, setting $l = 20$, and quantify the attractor density in each box $\rho(x,y)$ ($x,y = 1,\dots,l$) by measuring the fraction of the $N = 10^4$ recorded $x_1,y_1$ points that appear within each $(x,y)$ box (Fig.\ \ref{fig4}a). Regions of high density (yellow) represent areas in the $(x_1,y_1)$ plane that are frequently visited by the master oscillator, whereas regions of low density (blue) are rarely crossed by the oscillator. Denoting the maximum density by $\rho_{\rm max}$, we define the fractional density to be $\rho(x,y)/\rho_{\rm max}$, and mark the most (least) dense regions in the attractor, in which the fractional density exceeds $0.7$ (is beneath $0.2$) by red circles (triangles). To evaluate the level of local topological synchronization we measure the number of synchronization points $S(x,y)$ in each box and obtain the locally normalized SPP via $\sigma(x,y) = S(x,y)/\rho(x,y)$, capturing the probability of an attractor point in the $(x,y)$ box to be synchronized. Interestingly, we find that synchronization tends to avoid the high density regions (red circles), preferentially nucleating around the attractor regions that are scarcely visited by the master oscillator (red triangles, Fig.\ \ref{fig4}b-d).

\begin{figure}
\centering
\includegraphics[width=\columnwidth]{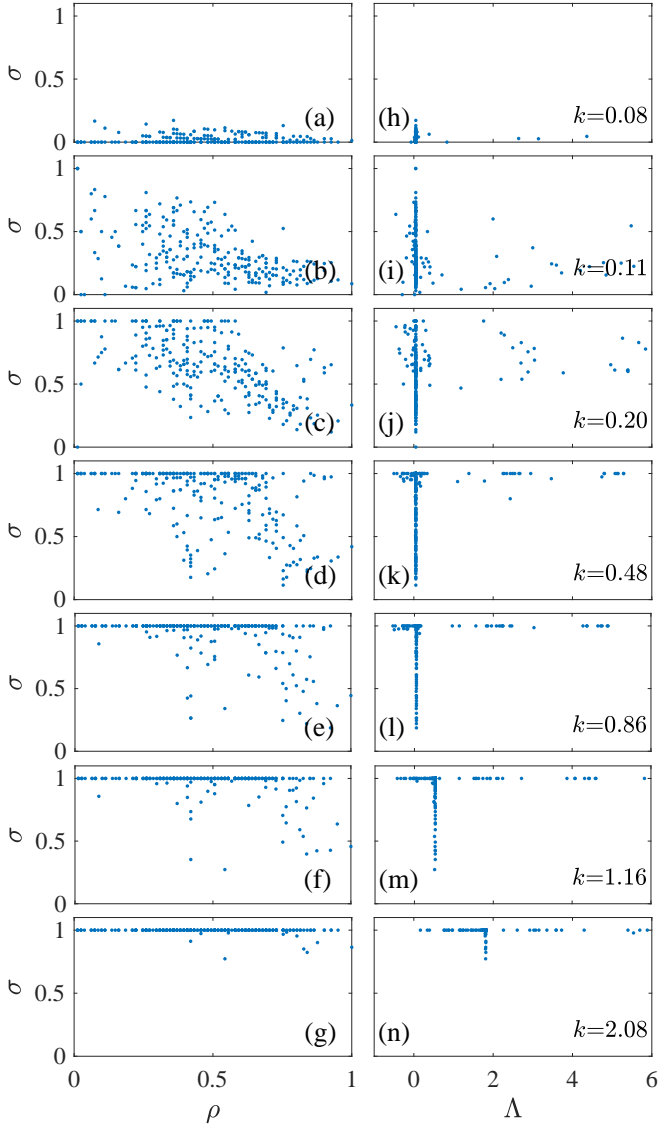}
\caption[]{(Color online). 
\textbf{The effect of density $\rho$ and local expansion rate $\Lambda$}.
(a) - (c) Normalized SPP $\sigma$ vs.\ the master attractor density $\rho$ as obtained for $k = 0.08, 0.11$ and $0.20$. Synchronization (large $\sigma$) begins in small $\rho$ regions of the attractor, gradually spreading towards large $\rho$. 
(d) - (g) For larger values of $k$, complete synchronization is gradually reached and eventually $\sigma \rightarrow 1$ throughout all points of the attractor. 
(h) - (n) $\sigma$ vs.\ $\Lambda$, the expansion rate at the slave image of all synchronization points. Initially, synchronization points show a strong preference for areas characterized by small $\Lambda$. Increasing $k$ the effect is twofold: the peak of the distribution shifts to the right hand side, \textit{i.e.} larger $\Lambda$, and more points begin to distribute uniformly, capturing the expanding coverage of the topological synchronization.}
\label{fig5}
\end{figure}

To systematically quantify this preference we show in Fig.\ \ref{fig5}a the normalized SPP, $\sigma(x,y)$, vs.\ the attractor density, $\rho(x,y)$, for $k = 0.08$, representing the first stages of topological synchronization. Indeed, we find that synchronization appears first only in the limit of small $\rho$, tending to zero as $\rho$ is increased. This trend is sustained for larger $k$ values (Fig.\ \ref{fig5}b-g), with $\sigma$ gradually advancing from the sparse regions towards the denser ones, until, once $k > k_{\rm CS} \sim 2$, we have $\sigma \rightarrow 1$ for all $\rho$, capturing complete synchronization (Fig.\ \ref{fig5}g).

The difficulty in synchronizing chaotic systems is rooted in their sensitivity to initial conditions, as quantified by the expansion rate, capturing the divergence of infinitesimally close trajectories. Therefore, we expect that local synchronization will appear first in those areas where the slave attractor is characterized by a low expansion rate. To test this we measured the maximum eigenvalue $\Lambda$ of the local Jacobian matrix, Eq.\ (\ref{jacobian}) at all points in the slave attractor and compared their expansion rate with the normalized SPP, $\sigma$. We find, in Fig.\ \ref{fig5}h-m, that indeed the majority of synchronization points (large $\sigma$) condense around slave attractor regions with low expansion rates (small $\Lambda$). This trend is consistently observed until $k \approx k_{\rm CS}$, where complete synchronization emerges and $\sigma \rightarrow 1$ almost homogeneously throughout the entire range of $\Lambda$ values (Fig.\ \ref{fig5}n).

\begin{figure}
\centering
\includegraphics[width=0.65\columnwidth]{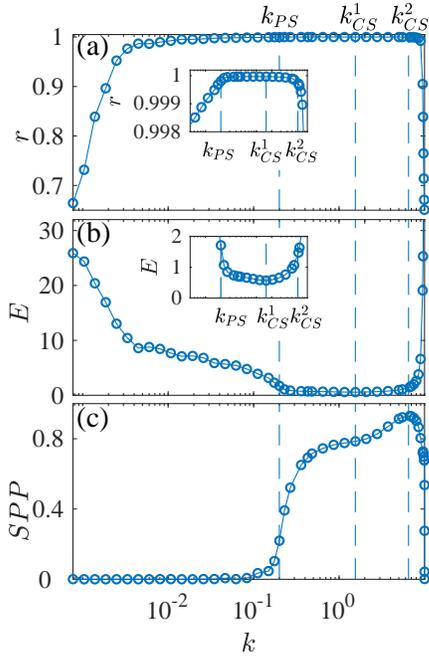}
\vspace{-4mm}
\caption[]{(Color online).
\textbf{Microscopic synchronization in a class III system}.
We constructed a R\"{o}ssler master slave system coupled only through the $x$ variable and observed the transitions in and out of synchronization.
(a) Phase order parameter $r$ vs.\ coupling strength $k$. Phase synchronization ($r \rightarrow 1$) occurs between $k > k_{\rm PS} \sim 0.2$ and begins to deteriorate at $k > k^{2}_{\rm CS} \sim 6.5$ (vertical dashed lines). The inset shows the behavior of $r$ in the vicinity of these transitions. 
(b) Synchronization error $E$ vs.\ the coupling strength $k$. We obtain a window of approximate complete synchronization, in which $E \ll 1$, between $k^{1}_{\rm CS} \sim 1.55$ and $k^{2}_{\rm CS} \sim 6.5$ (Vertical dashed lines, see also inset). 
(c) As in Fig.\ \ref{fig1}, SPP vs.\ $k$ allows us to observe the microscopic path to synchronization. As expected, SPP grows gradually with $k$ and then declines after the transition to the desynchronized state at $k > k^{2}_{\rm CS}$. Note the SPP never reaches unity, as, indeed, this system never fully exhibits complete synchronization (see inset in (b), where it is shown that $E$ never fully vanishes). 
\label{fig6}}
\vspace{-3mm}
\end{figure}

To further test our microscopic view of synchronization we consider a different form of master-slave coupling, in which the master and slave are coupled through a single variable $x$. This can be achieved by substituting the coupling term in Eq.\ (\ref{ecuaciones}) with $k(x_1 - x_2)$, a coupling applied uniquely to the $x$ variable. This seemingly minor change is rather essential, turning the coupled oscillators into a class III system, in which synchronization is confined to a finite interval in $k$ \cite{Boccaletti2006, Huang2009}. Indeed, now the system features a first transition at $k = k^1_{\rm CS}$, in which it enters complete synchronization, followed by a second transition at $k = k^2_{\rm CS} > k^1_{\rm CS}$, in which it begins to de-synchronize (Fig.\ \ref{fig6}a,b). Hence, synchronization is limited to the range $k^1_{\rm CS} < k < k^2_{\rm CS}$, allowing us to test our microscopic framework under a broader setting, which includes also the transition \textit{out} of synchronization.

\begin{figure}
\centering
\includegraphics[width=\columnwidth]{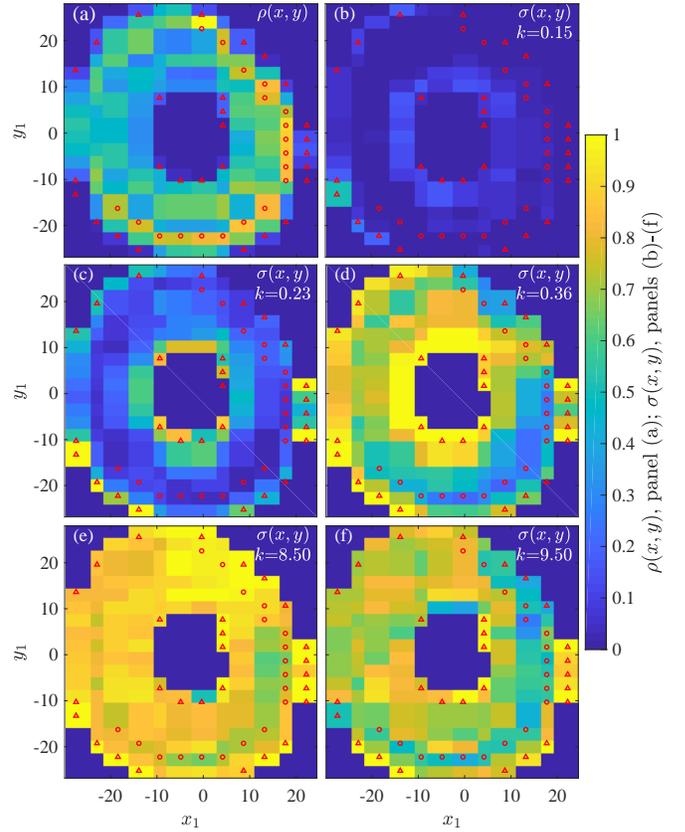}
\caption[]{(Color online).
\textbf{Synchronization focal points in a class III system}.
Coupled only through the $x$ coordinate, the system undergoes synchronization in a limited $k$ interval. Here we track the build up of local synchronization as the system enters and exits this interval. 
(a) Density $\rho(x,y)$ of points in the $(x_1,y_1)$ projection of the master attractor. As in Fig. 4 boxes with high density (above $70\%$ of $\max (\rho(x,y))$) are marked by red circles; boxes with low density (below $20\%$ of $\max(\rho(x,y))$) are marked by red triangles. 
(b) - (d) The normalized local SPP $\sigma(x,y)$ under increasing values of the coupling strength $k$. Early traces of synchronization (tending to yellow) favor the low density regions of the attractor, as observed by $\sigma(x,y)$'s consistent avoidance of the red circles and its tendency towards the red triangles. (e) - (f) At $k > k^{2}_{\rm CS} \sim 6.5$ the system exits the synchronization window. Once again, we find that this process occurs unevenly throughout the attractor, with regions of low density (triangles), again, being the last to de-synchronize.}
\label{fig7}
\end{figure}

\begin{figure}
\centering
\includegraphics[width=\columnwidth]{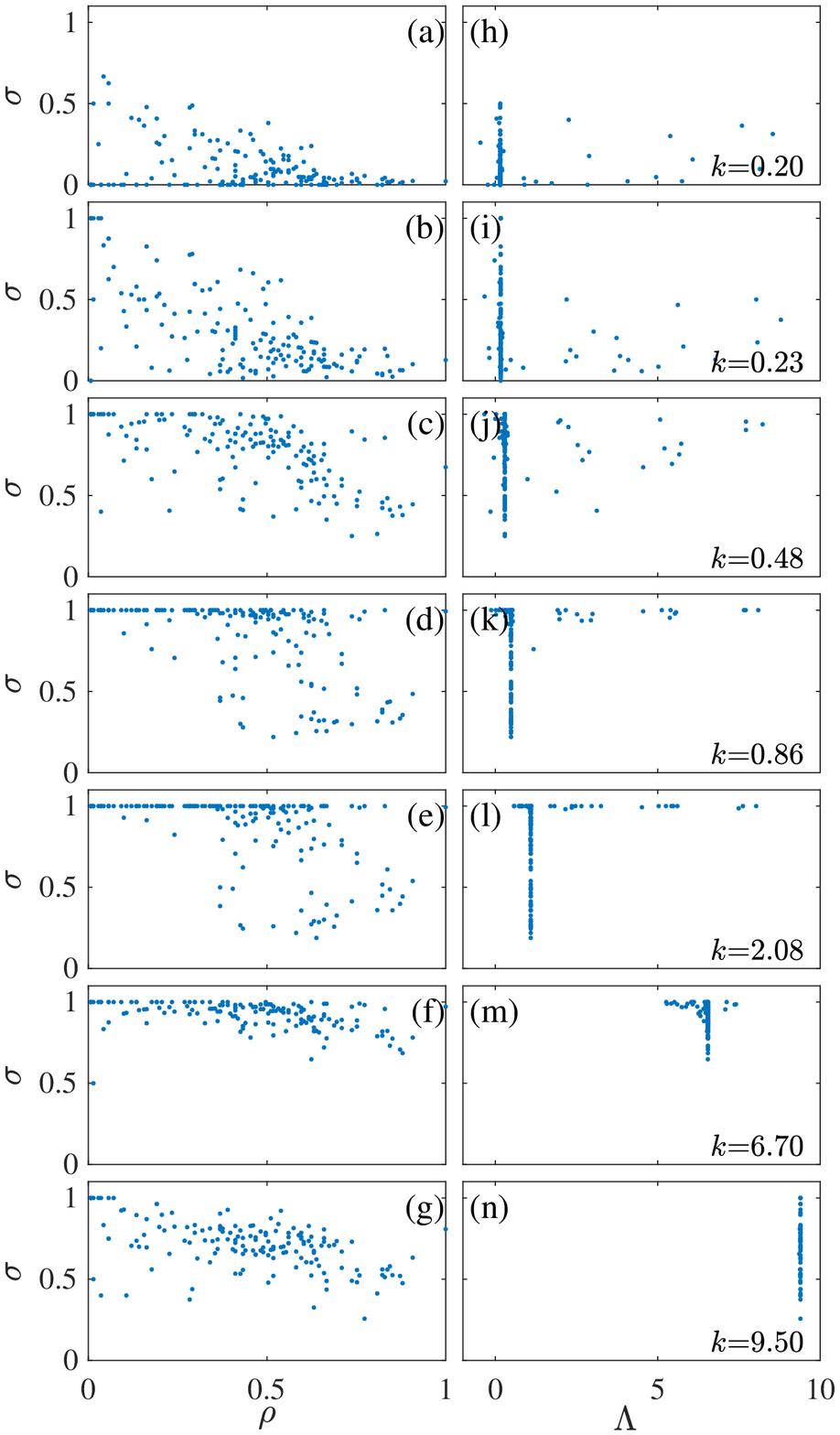}
\caption[]{(Color online). 
\textbf{The effect of $\rho$ and $\Lambda$ in a class III system}.
(a) - (c) Normalized SPP $\sigma$ vs.\ the master attractor density $\rho$ as obtained for $k = 0.2, 0.23$ and $0.48$. Synchronization (large $\sigma$) is initiated at low density regions of the attractor, gradually spreading towards large $\rho$. 
(d) - (e) For larger values of $k$, complete synchronization is gradually reached and $\sigma \rightarrow \sim 1$ throughout all points of the attractor. 
(f) - (g) This density-synchronization correlation is sustained as the system exits the synchronization window ($k > k^{2}_{\rm CS} \sim 6.5$), as the high density regions lose synchronization first.
(h) - (l) $\sigma$ vs.\ $\Lambda$, the expansion rate at the slave image of all synchronization points. Synchronization points show a strong preference towards areas characterized by small $\Lambda$. As in class II, increasing $k$, we find that the peak of the distribution shifts towards higher $\Lambda$, and more and more points tend to spread out uniformly. 
(m) - (n) For large $k$, de-synchronization begins, but, this time, the final traces of local synchronization are condensed around the large $\Lambda$ regions.  
}
\label{fig8}
\end{figure}

We find, also for this class III system that the SPP index captures the microscopic emergence of synchronization, exhibiting non-vanishing levels of local synchronization already before entering the complete synchronization window $k^1_{\rm CS} < k < k^2_{\rm CS}$ (Fig.\ \ref{fig6}c). As expected SPP begins to decline shortly after $k > k^2_{\rm CS}$, as synchronization begins to deteriorate. Interestingly, we observe that SPP$< 1$ throughout the entire range of coupling strengths, indicating that this system never attains perfect synchronization. Indeed, the inset in Fig.\ \ref{fig6}b shows that the order parameter $E$ never reaches the $E = 0$ mark, a slight deviation that has a clear microscopic fingerprint in SPP. As above, we find that synchronization is initiated at low density regions of the attractor (Fig.\ \ref{fig7} and Fig.\ \ref{fig8}a - g). This preference is sustained both upon entering the synchronized regime and upon exiting it at $k > k_{\rm CS}^2$. The picture is richer with respect to the local expansion rates $\Lambda$: the onset of synchronization is rooted in the low expansion regions of the slave attractor (Fig.\ \ref{fig8}h - l), consistent with our observations in Class II (Fig.\ \ref{fig5}h - n). However, as the system exits the synchronization window, under large $k$, the final traces of local synchronization are observed at the high expansion regions (Fig.\ \ref{fig8}m,n).

these correlations between synchronization points and density/expansion rates, observed in both systems tested, are of particular relevance, as they provide useful criteria to foresee the conditions under which one can predict the state of one system by measuring that of the other. For instance, consider the phase synchronization regime, in which phases are coordinated, but the amplitudes of the two chaotic systems remain globally uncorrelated. It would seem, due to the lack of amplitude correlations, that one cannot gain information on, \textit{e.g.}, $\mathbf{x}_2$ from measurements of $\mathbf{x}_1$. Our framework and analysis, however, uncovered that while \textit{global} synchronization is lacking, \textit{local} synchronization points may already be present, allowing prediction from $\mathbf{x}_1$ to $\mathbf{x}_2$ in certain areas of the oscillators' trajectory. For example, we have shown that around the transition to phase synchronization SPP $\sim 30\%$ (Fig.\ \ref{fig1}c), indicating that such mapping from $\mathbf{x}_1$ to $\mathbf{x}_2$ indeed exists in selected regions of the attractor. Equipped with knowledge on the density (of the master attractor) or the local expansion rates (of the slave) one can identify \textit{a priori} the most likely regions in the two attractors, where such prediction is potentially possible.

In this paper, we limited explicitly ourselves to the problem of complete synchronization in order to demonstrate the continuous microscopic process of topological synchronization taking place already in small coupling strengths. Complete synchronization defined here as the state wherein the two systems evolve quasi-identically, i.e. with negligible synchronization errors (since a formal synchronization manifold does not exist for non-identical systems). In this microscopic process the system might be also in lag synchronization before it will reach complete synchronization. Microscopically, lag synchronization will just correspond to some small region where SPP is still consistently smaller than 1, but yet consistently larger than 0.
Complementing the well-established macroscopic and meso-scopic views, we have exposed here the microscopic onset of synchronization, as a gradual process of \textit{topological adjustment} observed by the two coupled chaotic systems. This view helped us demonstrate the presence of synchronization points even in the absence of global synchronization and show the road to complete synchronization as a continuous \textit{Topological synchronization} between the two attractors. Following the track of almost all research into synchronization of chaotic systems, our study focused on slightly mismatched oscillators coupled via diffusive coupling. Its insights, however, can be readily expanded to other systems and to different types of coupling functions. However, while our basic findings, pertaining to the microscopic topological convergence are likely universal, the specific path towards synchronization, \textit{i.e.} nucleating in regions of small $\Lambda$ and gradually spreading to areas of large $\Lambda$, may depend on the specific choice of coupling. Therefore, our results prompt further investigation into different types of coupling functions and their corresponding unique patterns of local synchronization. 

\vspace{10mm}
\noindent
\textbf{Acknowledgements}. 
The authors would like to thank Ashok Vaish and Ricardo Guti\'errez for their continuous support.
C.H. wishes to thank the Planning and Budgeting Committee (PBC) of the Council for Higher Education, Israel, for support. 
This work was supported by the US National Science Foundation - CRISP Award Number: 1735505 and by the Ministerio de Econom\'ia y Competitividad of Spain (projects FIS2013-41057-P and FIS2017-84151-P).

\end{document}